\documentclass[12pt]{article}
\usepackage{amsmath,amssymb}

\makeatletter \@addtoreset{equation}{section} \makeatother

\newread\epsffilein    
\newif\ifepsffileok    
\newif\ifepsfbbfound   
\newif\ifepsfverbose   
\newif\ifepsfdraft     
\newdimen\epsfxsize    
\newdimen\epsfysize    
\newdimen\epsftsize    
\newdimen\epsfrsize    
\newdimen\epsftmp      
\newdimen\pspoints     
\def\epsfbox#1{\global\def\epsfllx{72}\global\def\epsflly{72}%
   \global\def\epsfurx{540}\global\def\epsfury{720}%
   \def\lbracket{[}\def\testit{#1}\ifx\testit\lbracket
   \let\next=\epsfgetlitbb\else\let\next=\epsfnormal\fi\next{#1}}%
\def\epsfgetlitbb#1#2 #3 #4 #5]#6{\epsfgrab #2 #3 #4 #5 .\\%
   \epsfsetgraph{#6}}%
\def\epsfnormal#1{\epsfgetbb{#1}\epsfsetgraph{#1}}%
\def\epsfgetbb#1{%

\openin\epsffilein=#1 \ifeof\epsffilein\errmessage{I couldn't open
#1, will ignore it}\else

   {\epsffileoktrue \chardef\other=12
    \def\do##1{\catcode`##1=\other}\dospecials \catcode`\ =10
    \loop
       \read\epsffilein to \epsffileline
       \ifeof\epsffilein\epsffileokfalse\else

          \expandafter\epsfaux\epsffileline:. \\%
       \fi
   \ifepsffileok\repeat
   \ifepsfbbfound\else
    \ifepsfverbose\message{No bounding box comment in #1;
                                             using defaults}\fi\fi
   }\closein\epsffilein\fi}%

\def\epsfclipoff{\def\epsfclipstring{\ifepsfdraft\space clip\fi}}%
\epsfclipoff
\def\epsfsetgraph#1{%
   \epsfrsize=\epsfury\pspoints
   \advance\epsfrsize by-\epsflly\pspoints
   \epsftsize=\epsfurx\pspoints
   \advance\epsftsize by-\epsfllx\pspoints

   \epsfxsize\epsfsize\epsftsize\epsfrsize
   \ifnum\epsfxsize=0 \ifnum\epsfysize=0
      \epsfxsize=\epsftsize \epsfysize=\epsfrsize
      \epsfrsize=0pt

     \else\epsftmp=\epsftsize \divide\epsftmp\epsfrsize
       \epsfxsize=\epsfysize \multiply\epsfxsize\epsftmp
       \multiply\epsftmp\epsfrsize \advance\epsftsize-\epsftmp
       \epsftmp=\epsfysize
       \loop \advance\epsftsize\epsftsize \divide\epsftmp 2
       \ifnum\epsftmp>0
          \ifnum\epsftsize<\epsfrsize\else
             \advance\epsftsize-\epsfrsize \advance\epsfxsize\epsftmp \fi
       \repeat
       \epsfrsize=0pt
     \fi
   \else \ifnum\epsfysize=0
     \epsftmp=\epsfrsize \divide\epsftmp\epsftsize
     \epsfysize=\epsfxsize \multiply\epsfysize\epsftmp
     \multiply\epsftmp\epsftsize \advance\epsfrsize-\epsftmp
     \epsftmp=\epsfxsize
     \loop \advance\epsfrsize\epsfrsize \divide\epsftmp 2
     \ifnum\epsftmp>0
        \ifnum\epsfrsize<\epsftsize\else
           \advance\epsfrsize-\epsftsize \advance\epsfysize\epsftmp \fi
     \repeat
     \epsfrsize=0pt
    \else
     \epsfrsize=\epsfysize
    \fi
   \fi

   \ifepsfverbose\message{#1: width=\the\epsfxsize, height=\the\epsfysize}\fi
   \epsftmp=10\epsfxsize \divide\epsftmp\pspoints
   \vbox to\epsfysize{\vfil\hbox to\epsfxsize{%
      \ifnum\epsfrsize=0\relax
        \includegraphics{\ifepsfdraft}%
      \else
        \epsfrsize=10\epsfysize \divide\epsfrsize\pspoints
        \includegraphics{\ifepsfdraft}%
      \fi
      \hfil}}%
\global\epsfxsize=0pt\global\epsfysize=0pt}%

{\catcode`\%=12 \global\let\epsfpercent=

\long\def\epsfaux#1#2:#3\\{\ifx#1\epsfpercent
   \def\testit{#2}\ifx\testit\epsfbblit
      \epsfgrab #3 . . . \\%
      \epsffileokfalse
      \global\epsfbbfoundtrue
   \fi\else\ifx#1\par\else\epsffileokfalse\fi\fi}%

\def\epsfempty{}%
\def\epsfgrab #1 #2 #3 #4 #5\\{%
\global\def\epsfllx{#1}\ifx\epsfllx\epsfempty
      \epsfgrab #2 #3 #4 #5 .\\\else
   \global\def\epsflly{#2}%
   \global\def\epsfurx{#3}\global\def\epsfury{#4}\fi}%

\def\epsfsize#1#2{\epsfxsize}

\let\epsffile=\epsfbox

\textwidth=6.0in
\hoffset=-.55in \textheight=9in \voffset=-.8in

\def\crampest{\medmuskip = 1mu plus 1mu minus 1mu}

\def\nn{\nonumber}

\let\bm=\bibitem

\def\be{\begin{equation}}
\def\ee{\end{equation}}
\def\ba{\begin{array}}
\def\ea{\end{array}}
\def\ft#1#2{{\textstyle{\frac{\scriptstyle #1}{\scriptstyle #2}}}}
\def\fft#1#2{\frac{#1}{#2}}

\def\sst#1{{\scriptscriptstyle #1}}

\def\td{\tilde}
\def\wtd{\widetilde}

\def\dalemb#1#2{{\vbox{\hrule height .#2pt
        \hbox{\vrule width.#2pt height#1pt \kern#1pt
                \vrule width.#2pt}
        \hrule height.#2pt}}}
\def\square{\mathord{\dalemb{6.8}{7}\hbox{\hskip1pt}}}

\newcommand{\hoch}[1]{$\, ^{#1}$}
\newcommand{\bea}{\begin{eqnarray}}
\newcommand{\eea}{\end{eqnarray}}

\def\0{{\sst{(0)}}}
\def\1{{\sst{(1)}}}
\def\2{{\sst{(2)}}}
\def\3{{\sst{(3)}}}
\def\4{{\sst{(4)}}}
\def\5{{\sst{(5)}}}
\def\6{{\sst{(6)}}}
\def\7{{\sst{(7)}}}
\def\8{{\sst{(8)}}}

\def\im{{{\rm i}}}

\def\ep{{\epsilon}}

\def\R{\rlap{\rm I}\mkern3mu{\rm R}}

\def\R{\rlap{\rm I}\mkern3mu{\rm R}}

\def\R{{{\mathbb R}}}

\def\CP{{{\mathbb C}{\mathbb P}}}

\def\Z{{{\mathbb Z}}}

\thispagestyle{empty}

\begin{document}
\begin{flushright}
MIFP-06-33\ \ \ \ UPR-1172-T\\
{\bf hep-th/0701082}\\
January\  2007
\end{flushright}

\vspace{10pt}
\begin{center}

{\Large {\bf Resolved Calabi-Yau Cones and Flows\\ from $L^{abc}$
Superconformal Field Theories}}

\vspace{20pt}

W. Chen$^{\dagger 1}$, M. Cveti\v{c}$^{\ddagger 2}$, H. L\"u$^{\dagger
1}$, C.N. Pope$^{\dagger 1}$ and J.F. V\'azquez-Poritz$^{\dagger 1}$

\vspace{20pt}

{\hoch{\dagger}\it George P. \&  Cynthia W. Mitchell Institute
for Fundamental Physics\\
Texas A\&M University, College Station, TX 77843-4242, USA}

\vspace{10pt}

{\hoch{\ddagger}\it Department of Physics \& Astronomy\\
University of Pennsylvania, Philadelphia, PA 19104-6396, USA}

\vspace{40pt}

\underline{ABSTRACT}
\end{center}

We discuss D3-branes on cohomogeneity-three resolved Calabi-Yau cones
over $L^{abc}$ spaces, for which a 2-cycle or 4-cycle has been blown
up. In terms of the dual quiver gauge theory, this corresponds to
motion along the non-mesonic, or baryonic, directions in the moduli
space of vacua.  In particular, a dimension-two and/or dimension-six
scalar operator gets a vacuum expectation value. These resolved cones
support various harmonic $(2,1)$-forms which reduce the ranks of some
of the gauge groups either by a Seiberg duality cascade or by
Higgsing. We also discuss higher-dimensional resolved Calabi-Yau
cones.  In particular, we obtain square-integrable $(2,2)$-forms for
eight-dimensional cohomogeneity-four Calabi-Yau metrics.

{\vfill\leftline{}\vfill \vskip 10pt \footnoterule {\footnotesize
{\footnotesize
\hoch{1} Research supported in part by DOE grant
DE-FG03-95ER40917.}\vskip 2pt
\hoch{2} Research supported in part by DOE grant
DE-FG02-95ER40893, NSF grant INTO3-24081, $\phantom{xxxxi}$ and the
 Fay R. and Eugene L. Langberg Chair.}\vskip 2pt}

\newpage
\tableofcontents
\addtocontents{toc}{\protect\setcounter{tocdepth}{2}}
\newpage

\section{Introduction}

The AdS/CFT correspondence relates type IIB string theory on
AdS$_5\times S^5$ to four-dimensional ${\cal N}=4$ $U(N)$
superconformal Yang-Mills theory \cite{ads1,ads2,ads3}. More
generally, type IIB string theory on AdS$_5\times X^5$, where $X^5$ is
an Einstein-Sasaki space such as $T^{1,1}$, $Y^{pq}$ \cite{Ypq1,Ypq2}
or $L^{abc}$ \cite{Lpqr1,Lpqr2}, corresponds to an ${\cal N}=1$
superconformal quiver gauge theory. The dual gauge theories have been
identified in \cite{T11} for $T^{1,1}$, in \cite{martelli,YpqCFT} for
$Y^{pq}$ and in \cite{LpqrCFT1,LpqrCFT2,LpqrCFT3} for $L^{abc}$.

There is a prescription for mapping perturbations of the supergravity
background to operators in the dual gauge theory \cite{ads2,ads3}. In
particular, motion in the K\"ahler moduli space of the Calabi-Yau cone
over the Einstein-Sasaki space corresponds to giving vacuum
expectation values (vevs) to the fundamental fields, such that only
non-mesonic operators get vevs. This is because the mesonic directions
of the full moduli space correspond to the motion of the D3-branes in
the Calabi-Yau space whereas the non-mesonic, or baryonic, directions are
associated with either deformations of the geometry or turning on
$B$-fields. This has been studied for a blown-up 2-cycle in the
resolved conifold in \cite{resolvedT11}, as well as for a blown-up
4-cycle in the resolved conifold \cite{pandotseytlin1}, $Y^{pq}$ cones
\cite{pal}, $L^{abc}$ cones \cite{sfetsos} and general Calabi-Yau
cones \cite{leo}. All of these resolved Calabi-Yau cones with blown-up
4-cycles follow the general construction given in
\cite{bergery,pagepope}.

In this paper, we shall apply the state/operator correspondence to a
general class of resolved Calabi-Yau cones over $L^{abc}$ with a
blown-up 2-cycle or 4-cycle. These metrics can be obtained from the
Euclideanization of the BPS limit of the six-dimensional Kerr-NUT-AdS
solutions \cite{nutbh1,nutbh2}.\footnote{This is the even-dimensional
analog of the relation between the Einstein-Sasaki spaces constructed
in \cite{lupopo} and odd-dimensional BPS Kerr-NUT-AdS solutions.} In
particular, blowing up a 2-cycle or 4-cycle corresponds to giving a
vev to a real dimension-two and/or six scalar operator. Although
cycles are being blown up, in all but two cases there remain
singularities \cite{oota,lupores}.  However, there is a countably
infinite subset of cases where there is an ALE singularity, on which
perturbative string dynamics is well-defined.  Some of these cases
were studied in \cite{leo}.  While adding a large number of D3-branes
that are uniformly distributed, or ``smeared", on the blown-up cycle ends 
up inducing a power-law singularity at short distance\footnote{This 
singularity is due to the smearing of the D3-brane charge on the blown-up 
cycle. A completely non-singular solution with D3-branes stacked at a single 
point on the resolved conifold has been constructed \cite{murugan}.}, the
resulting backgrounds can nevertheless be reliably used to describe
perturbations around the UV conformal fixed point of the quiver gauge
theories. Close to the UV fixed point, blowing up a 2-cycle on the
$L^{abc}$ cone corresponds to giving a vev to an operator that is
analogous to the case of the resolved conifold. Therefore, we shall
refer to these spaces as resolved cones, though it should be
understood that there are still orbifold-type singularities.

The supergravity background can also be perturbed by adding a harmonic
3-form which lives on the Calabi-Yau metrics. If this is a pure
$(2,1)$-form then supersymmetry will be preserved. Furthermore, if
this form carries nontrivial flux then it corresponds to D5-branes
wrapped on a 2-cycle in the Calabi-Yau space. The introduction of
these fractional D3-branes eliminates the conformal fixed point in the
UV limit of the quiver gauge theory. The theory undergoes a Seiberg
duality cascade and the ranks of some of the gauge groups are reduced
with decreasing energy scale. The supergravity solutions corresponding
to fractional branes have been constructed for the cones over $T^{11}$
\cite{frac1,frac2}, $Y^{pq}$ \cite{fracYpq} and $L^{abc}$ spaces
\cite{martellisparks,fracLpqr}. Fractional branes have also been
considered for Calabi-Yau spaces with blown-up cycles, such as the
deformed conifold \cite{ks}, resolved conifold \cite{pandotseytlin2}
and regularized conifold \cite{pandotseytlin1}, as well as the
resolved $Y^{pq}$ cones with blown-up 4-cycles \cite{leo}.  We shall
also consider continuous families of 3-forms that do not have
nontrivial flux. In this case, there remains a conformal fixed point
in the UV limit of the field theory. It has been proposed that the
ranks of some of the gauge groups are reduced with decreasing energy
scale via the Higgs mechanism \cite{aharony}.

Since the $L^{abc}$ spaces have cohomogeneity two, the form fields
constructed on the corresponding Calabi-Yau spaces will generally have
nontrivial dependence on the radial direction as well as the two
non-azimuthal coordinates of $L^{abc}$. In addition, these forms
generally break the $U(1)_R\times U(1)\times U(1)$ global symmetry
group of the theory down to a $U(1)\times U(1)$ symmetry group which,
in particular, breaks the R-symmetry. However, this is done in such a
way that the theory preserves ${\cal N}=1$ supersymmetry.

The various perturbations of the AdS$_5\times L^{abc}$ supergravity
background that will be discussed are shown in Figure
1. These perturbations, which can be superimposed with one another, 
correspond to continuous families of Renormalization Group (RG) flows 
from the UV superconformal fixed point of the quiver gauge theory. 
 
\begin{figure}[ht]
   \epsfxsize=3.7in \centerline{\epsffile{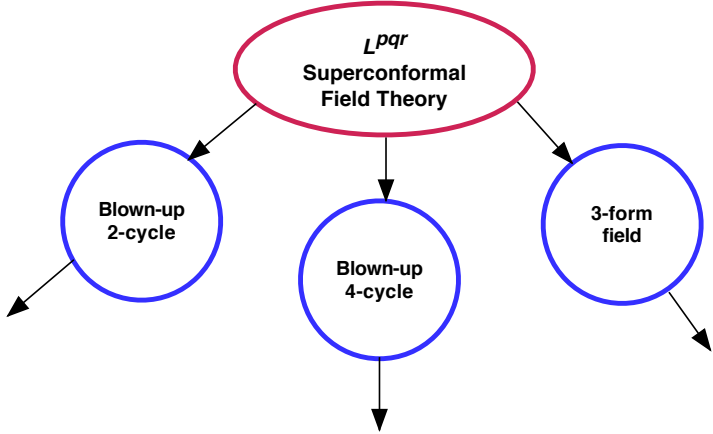}}
   \caption[FIG. \arabic{figure}.]{\footnotesize{RG flows from the
   superconformal fixed point of the $L^{abc}$ quiver gauge theory
   correspond to various deformations of the supergravity
   background.}}
   \label{figure1}
\end{figure}

The paper is organized as follows. In section 2, we discuss the
geometry of the resolved Calabi-Yau cones over the $L^{abc}$ spaces. A
subset of these are the resolved cones over $Y^{pq}$ and their various
limits. We find various harmonic $(2,1)$-forms on these metrics, some
of which carry nontrivial flux and some of which do not. In section 3,
we apply some of our results to the AdS/CFT correspondence. In
particular, we relate the perturbations of the AdS$_5\times L^{abc}$
background to various flows from the UV conformal fixed point of the
dual quiver gauge theory. In section 4, we consider eight-dimensional
resolved cones over $L^{pqrs}$ and the various harmonic forms that
live on them. In section 5, we carry out the corresponding analysis
for the higher-dimensional resolved cones.  Lastly, conclusions are
presented in section 6.

\section{Six-dimensional resolved Calabi-Yau cones}

Although the $L^{abc}$ spaces themselves are non-singular for
appropriately chosen integers $p$, $q$, $r$ \cite{Lpqr1,Lpqr2}, the
cones over these spaces have a power-law singularity at their apex. In
the case of the cone over $T^{1,1}$, this singularity can be smoothed
out in two different ways \cite{candelas}. Firstly, one can blow up a
3-cycle, which corresponds to a complex deformation. The resulting
deformed conifold has been crucial for the construction of a
well-behaved supergravity dual of the IR region of the gauge theory,
providing a geometrical description of confinement \cite{ks}.

One might hope that a similar resolution procedure could be performed
on other $L^{abc}$ cones.  Although a first-order deformation of the
complex structure of $Y^{pq}$ cones has been found in
\cite{burrington}, there exists an obstruction to finding the complex
deformations beyond first order \cite{altmann1,altmann2}. There is
also evidence from the field theory side that such deformations will
break supersymmetry for the $Y^{pq}$ cones 
\cite{nosusy1,nosusy2,nosusy3,forcella} as well as for a large class of 
$L^{abc}$ cones \cite{forcella}. Nevertheless,
there are $L^{abc}$ cones which allow for complex structure
deformations \cite{altmann1,altmann2}, which can be understood from
the corresponding toric diagrams \cite{LpqrCFT2}.\footnote{We thank
Angel Uranga for discussions on this point.}  However, the explicit
metrics for these deformed $L^{abc}$ cones, let alone the solutions
for D3-branes on these cones, are not known.

\begin{figure}[ht]
   \epsfxsize=1.2in \centerline{\epsffile{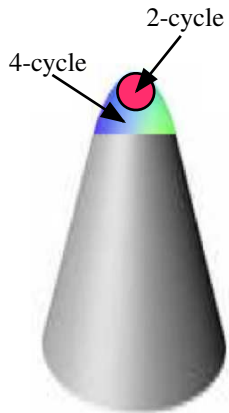}}
   \caption[FIG. \arabic{figure}.]{\footnotesize{A 4-cycle within the
   base space of a cone over $L^{abc}$} can be blown up. Within this
   4-cycle lies a 2-cycle. The volumes of these two cycles correspond
   to two independent K\"ahler moduli.}
   \label{figure2}
\end{figure}

The second way in which the $T^{1,1}$ cone can be rendered regular is
by blowing up a 2-cycle \cite{candelas}. Also, for the case of a cone
over $T^{1,1}/\Z_2$, the singularity can be resolved by blowing up a
4-cycle. Both of these resolutions are examples of K\"ahler
deformations which, as we shall see shortly, can also be performed on
the $L^{abc}$ cones $C(L^{abc})$. Moreover, the 2-cycle actually lives
within the 4-cycle, as illustrated in Figure 2. This means that there
are two K\"ahler moduli associated with the 4-cycle.  For certain
parameter choices, we can have the 4-cycle corresponds to the
Einstein-K\"ahler base space of $L^{abc}$, whose metric can be
obtained by taking a certain scaling limit of a Euclideanized form of
the Plebanski-Demianski metric \cite{martellisparks}. It is also
possible to have the volume of the 4-cycle vanishes, whilst keeping a
2-cycle blown up.

It has been found that the cone over $Y^{2,1}$ can be rendered
completely regular by blowing up an appropriate 4-cycle
\cite{oota}. However this, together with the resolved cones over
$T^{1,1}$ and $T^{1,1}/\Z_2$, constitute the only examples of
non-singular resolved cones over the $L^{abc}$ spaces
\cite{lupores}. Although we shall refer to these spaces as
``resolved'' $L^{abc}$ cones, there are generally orbifold-type
singularities remaining. In the limit of a vanishing 2-cycle, this can
be seen simply because at short distance the geometry becomes a direct
product of $\R^2$ and the four-dimensional Einstein-K\"{a}hler base
space of $L^{abc}$, which is itself an orbifold. Nevertheless, the
resolved cones over $L^{abc}$ can be embedded in ten dimensions to
give Ricci-flat backgrounds Mink$_4\times C(L^{abc})$, on which
perturbative string dynamics is well-defined. However, as we shall see
in section 3, the back-reaction of D3-branes leads to a power-law
singularity at short distance. This singularity is due to the fact that we are 
smearing the D3-branes on the blown-up cycle. For the case of the 
resolved conifold, it has been shown that if the D3-branes are stacked 
at a single point then the supergravity solution is completely regular 
\cite{murugan}.

\subsection{Resolved cones over $Y^{pq}$}

Before turning to resolved cones over the general cohomogeneity-two
$L^{abc}$ spaces, it is instructive first to consider the subset
involving the cohomogeneity-one $Y^{pq}$ spaces. The resolved cone
over $Y^{pq}$ has the metric \cite{nutbh1}
\be
ds_6^2=\fft{x+y}{4X} dx^2 + \fft{X}{x+y} (d\tau + \fft{y}{2\alpha}
\,\sigma_3)^2 +
\fft{x+y}{4Y} dy^2 + \fft{Y}{x+y} (d\tau -\fft{x}{2\alpha}
\, \sigma_3)^2
+\fft{ x\, y}{4\alpha}\, (\sigma_1^2 + \sigma_2^2)\,.
\label{Ypqconemetric}
\ee
where
\be
X=x(x+\alpha) - \fft{2\mu}{x}\,,\qquad
Y=y(\alpha-y) + \fft{2\nu}{y}\,,
\ee
and that
\be
\sigma_3=d\psi + \cos\theta\, d\phi\,,\qquad
\sigma_1^2 + \sigma_2^2 = d\theta^2 + \sin^2\theta\, d\phi^2\,.
\ee
It has been shown that the only completely regular examples are the
resolved cones over $T^{1,1}$, $T^{1,1}/\Z_2$ and $Y^{2,1}$
\cite{oota,lupores}.  We shall now consider various limits of the
metric (\ref{Ypqconemetric}).

\bigskip
\noindent\underline{Resolved conifold}
\bigskip

In order to reduce to a resolved cone over $T^{11}$ (or $T^{11}/\Z_2$),
we need
to select $\nu$ such that $Y(y)$ has a double root.  This happens when
$\nu=-\ft2{27}\alpha^3$. Making the coordinate redefinition
\be
y=\ft23\alpha + \epsilon\, \cos\td\theta\,,\qquad
\nu=-\ft2{27}\alpha^3 + \ft12\alpha \epsilon^2\,,\qquad
\tau=-\fft{2}{9\epsilon} \td \phi\,,\qquad
\sigma_3\rightarrow \sigma_3 + \fft{2\alpha}{3\epsilon} d\tau\,,
\ee
and setting the parameter $\epsilon$ to zero, we find that the metric
becomes
\be
ds_6^2=\fft{x+\ft23\alpha}{4X} dx^2 + \fft{X}{9(x+\ft23\alpha)}
(\sigma_3 + \cos\td\theta\,d\td\phi)^2 + \ft16 (x + \ft23\alpha)
(d\td\theta^2 + \sin^2\td\theta\, d\td\phi^2) + \ft16 x\, (\sigma_1^2 +
\sigma_2^2)\,.
\ee
If $\mu=0$, there is a blown-up $S^2$ and the solution describes the
resolved conifold \cite{candelas}.  If, on the other hand, $\alpha=0$,
then there is a blown-up $S^2\times S^2$ and the solution describes
the regularized conifold \cite{pandotseytlin1}.  In fact, it has been
shown that one can always blow up a 4-cycle on any cone over an
Einstein-Sasaki space \cite{bergery,pagepope}. We shall now take a
look at the analogous limits for the resolved cones over the $Y^{pq}$
spaces.

\bigskip
\noindent\underline{The $\alpha=0$ limit}
\bigskip

If we let $y\rightarrow \alpha y$, $\nu\rightarrow \alpha^3 \nu$ 
and then take $\alpha\rightarrow 0$, we obtain the limit
\be
ds^2=\fft{x}{4X} dx^2 + \fft{X}{x} (d\tau + \ft12 y\sigma_3)^2 +
x\Big[\fft{dy^2}{4Y} + Y\, \sigma_3^2 + \ft14 y (\sigma_1^2 + \sigma_2^2)
\Big]\,,
\ee
where
\be
X=x^2 - \fft{2\mu}{x}\,,\qquad
Y=y(1-y) + \fft{2\nu}{y}\,.
\ee
There is a single K\"ahler modulus, which corresponds to a blown-up
4-cycle with a volume parameterized by $\mu$. This is the analog of
the resolved cone for general $Y^{pq}$ spaces. However, unlike the
$T^{1,1}/\Z_2$ case, this metric has an orbifold-type singularity at
its apex, since the geometry reduces to the direct product of $\R^2$
and an Einstein-K\"ahler orbifold.

\bigskip
\noindent\underline{The $\mu=0$ limit: blowing up 2-cycles}
\bigskip

One can also consider the limit in which $\mu$ vanishes, in which
case $x$ runs from 0 to asymptotic $\infty$.   Near $x=0$,
we can express the metric as
\be ds^2=y \Big(dr^2 + \ft14 r^2(\sigma_3 + \fft{2}{y} d\tau)^2 +
\ft14r^2 (\sigma_1^2 + \sigma_2^2)+\fft{dy^2}{4Y}\Big) + Y(d\tau -
\ft12 r^2\sigma_3)^2\,, \ee
where $x=r^2$. At $r=0$ there is a collapsing 3-sphere, instead of a
circle as in the previous limit. There is a single K\"ahler modulus
corresponding to the volume of a blown-up 2-cycle, which is
parameterized by $\alpha$. However, unlike the analogous resolved
conifold for which there is a smooth 2-sphere, in general this 2-cycle
is a ``tear-drop" with a conical singularity.

\bigskip
\noindent\underline{Calabi-Yau structure}
\bigskip

The Calabi-Yau structure on the metric (\ref{Ypqconemetric}) is
given by a K\"ahler form $J$ and a holomorphic $(3,0)$-form
$G_{(3)}$. These can be expressed in the complex vielbein basis
\be
\epsilon^1 = e^1 + {\rm i}\, e^2\,,\qquad
\epsilon^2 = e^3 + {\rm i}\, e^4\,,\qquad
\epsilon^3 = e^5 + {\rm i}\, e^6\,,
\ee
where the vielbein is conveniently chosen to be
\bea
&&e^1=\sqrt{\frac{x+y}{4X}}\, dx\,,\quad e^2=\sqrt{\fft{X}{x+y}}\,
\Big( d\tau + \fft{y}{2\alpha}\,\sigma_3\Big)
\,,\quad e^3=\sqrt{\fft{x+y}{4Y}}\, dy\,,\nn\\
&&e^4=\sqrt{\fft{Y}{x+y}}\, \Big( d\tau - \fft{x}{2\alpha}\,\sigma_3\Big)
\,,\quad
 e^5=\sqrt{\fft{xy}{4\alpha}}\,\sigma_1\,,\quad 
e^6=\sqrt{\fft{xy}{4\alpha}}\,\sigma_2\,.\label{d6ypqviel}
\eea
The K\"ahler 2-form is then given by
\be
J=\ft{\rm i}2\, \epsilon^i\wedge\bar\epsilon^i\,,
\ee
and the complex self-dual harmonic $(3,0)$-form is given by
\be
G_\3=e^{-3{\rm i} \tau}\,\epsilon^1\wedge\epsilon^2\wedge \epsilon^3
\equiv W_\3 + {\rm i}\,{*W_\3}\,.\label{g3}
\ee

\bigskip
\noindent\underline{Harmonic $(2,1)$-forms}
\bigskip

We are interested in harmonic $(2,1)$-forms that live on the resolved
$Y^{pq}$ cones, since their presence preserves the minimal
supersymmetry of the theory. We find there exist the following five
such $(2,1)$-forms:
\crampest{
\bea
\Phi_1 &=& \fft{e^{-3{\rm i}\tau}}{x\, X}\,
\bar\epsilon_1\wedge\epsilon_2\wedge \epsilon_3\,,\quad
\Phi_2 = \fft{e^{-3{\rm i}\tau}}{y\, Y}\,
\bar\epsilon_2\wedge\epsilon_1\wedge \epsilon_3\,,\quad
\Phi_3 = \fft{e^{3{\rm i}\tau}}{x\, y\, X\, Y}
\bar\epsilon_3\wedge \epsilon_1\wedge \epsilon_2\,,\nn\\
\Phi_4&=&\fft{1}{x\,y\,\sqrt{x+y}} \Big( \fft{1}{x\sqrt{Y}}
\epsilon_2\wedge (\bar\epsilon_3\wedge\epsilon_3 - \bar\epsilon_1
\wedge\epsilon_1)
-\fft{1}{y\sqrt{X}} \epsilon_1\wedge
(\bar\epsilon_3\wedge\epsilon_3 - \bar\epsilon_2\wedge\epsilon_2)\Big)
\,,\nn\\
\Phi_5&=&\fft{1}{\sqrt{x+y}} \Big( \fft{1}{x^2\sqrt{Y}}
\epsilon_2\wedge (\bar\epsilon_3\wedge\epsilon_3 - \bar\epsilon_1
\wedge\epsilon_1)
+\fft{1}{y^2\sqrt{X}} \epsilon_1\wedge
(\bar\epsilon_3\wedge\epsilon_3 - \bar\epsilon_2\wedge\epsilon_2)\Big)
\,.\label{Ypqforms}
\eea
}
All of these forms have singularities at all distances $x$, for
certain values of $y$, except for $\Phi_1$, which has a singularity
only at small distance.  $\Phi_1$ has a rapid fall off at large
distance, such that it does not support nontrivial flux. On the other
hand, in the large-$x$ limit the last harmonic form behaves like
\bea
\Phi_5 &=& \ft14 \sigma_1\wedge\sigma_2 \wedge (
\sigma_3 +\fft{2}{y}\,d\tau) + \fft{1}{2y^2} \sigma_3\wedge
d\tau \wedge dy + \fft{1}{4x}\Big(-2 \sigma_1\wedge\sigma_2\wedge d\tau
\nn\\
&&\quad +
{\rm i}\, \Big[ (\fft{1}{y}\sigma_1\wedge\sigma_2 -
\fft{1}{y^2} \sigma_3\wedge dy)\wedge dx +
\fft{y}{Y} \sigma_1\wedge\sigma_2\wedge dy\Big]\Big)+
{\cal O}(\fft{1}{x^2}).
\eea
This indicates that this form does support nontrivial flux. In the
$\alpha=0$ limit, in which we have first rescaled $y\rightarrow \alpha
y$, $\Phi_4$ and $\Phi_5$ reduce to the same form. This form has a
singularity that is confined to small distance.

\subsection{Resolved cones over $L^{abc}$}

We now turn to the resolved cones over the general cohomogeneity-two
$L^{abc}$ spaces. The metric is given by \cite{nutbh2}
\bea
ds^2&=&\ft14 (u^2 dx^2 + v^2 dy^2 + w^2 dz^2) +
\fft{1}{u^2} (d\tau + (y+z) d\phi + y z\, d\psi)^2\nn\\
&& +
\fft{1}{v^2} (d\tau + (x+z) d\phi + x z\, d\psi)^2 +
\fft{1}{w^2} (d\tau + (x+y) d\phi + x y\, d\psi)^2,
\eea
where the functions $u,v,w$ are given by
\bea
&&u^2=\fft{(y-x)(z-x)}{X}\,,\qquad
v^2=\fft{(x-y)(z-y)}{Y}\,,\qquad
w^2=\fft{(x-z)(y-z)}{Z}\,,\nn\\
&&X=x(\alpha-x)(\beta-x) - 2M\,,\qquad
Y=y(\alpha-y)(\beta-y) - 2L_1\,,\nn\\
&&
Z=z(\alpha-z)(\beta-z) - 2L_2\,. \label{uvwXYZ}
\eea
Notice that the coordinates $x$, $y$ and $z$ appear in the metric on a
symmetrical footing. We shall choose $x$ to be the radial direction,
and $y$ and $z$ to be the non-azimuthal coordinates on the $L^{abc}$
level sets.  This reduces to the $Y^{pq}$ subset when $a=p-q$, $b=p+q$
and $c=d=p$.

\bigskip
\noindent\underline{Calabi-Yau structure}
\bigskip

The complex vielbein can be written as
\be
\epsilon^1 = e^1 + {\rm i}\, e^2\,,\qquad
\epsilon^2 = e^3 + {\rm i}\, e^4\,,\qquad
\epsilon^3 = e^5 + {\rm i}\, e^6\,,
\ee
in the vielbein basis
\bea &&e^1=\ft12 u\, dx\,,\qquad e^2=\fft{1}{u}(d\tau + (y+z) d\phi +
y z\,d\psi)\,,\nn\\ &&e^3=\ft12 v\, dy\,,\qquad e^4=\fft{1}{v}(d\tau +
(x+z) d\phi + x z\,d\psi)\,,\nn\\ &&e^5=\ft12 w\, dz\,,\qquad
e^6=\fft{1}{w}(d\tau + (x+y) d\phi + x y\,d\psi)\,.
\eea
Then the K\"ahler 2-form and complex self-dual harmonic $(3,0)$-form
are given by
\be
J=\fft{\rm i}{2} \bar\epsilon_i\wedge\epsilon_i\,,\qquad
G_\3=e^{\im\, \nu}\, \epsilon_1\wedge\epsilon_2\wedge\epsilon_3\,,
\ee
where 
\be
\nu= 3\tau + 2(\alpha+\beta)\phi + \alpha\beta\psi\,.
\ee

\bigskip
\noindent\underline{Harmonic $(2,1)$-forms}
\bigskip

There is a harmonic $(2,1)$-form given by
\be
\Psi_1 = \fft{e^{\im\, \nu}}{X}\, \bar\epsilon^1\wedge\epsilon^2
   \wedge\epsilon^3\,.\label{psi1}
\ee
Using this, one can then construct a general class of harmonic
$(2,1)$-forms
\be
\Phi_1= f(\gamma)\, \Psi_1\,,
\ee
for any function $f$ so long as $d\gamma\wedge\Psi_1=0$. This
orthogonality condition is obeyed by
\be
\gamma= \fft{Y Z}{X}\, e^{\im\, 2\nu}\,,
\ee
as can be seen by calculating its exterior derivative:
\be
d\gamma= \fft{2\gamma}{(x-y)(y-z)(z-x)}\, 
 \Big( u(y-z) X'\, \bar\epsilon^1 - v(z-x) Y'\, \epsilon^2 -
    w (x-y) Z'\, \epsilon^3\Big)\,.
\ee
We can consider the special case for which 
\be
\Phi_1=\fft{(YZ)^{\delta}}{X^{\delta+1}} e^{\im(2\delta+1)\nu}\,
\bar\epsilon^1\wedge\epsilon^2\wedge\epsilon^3\,,
\ee
where $\delta$ is a continuous parameter. Due to the $\nu$ dependence,
this field only preserves $U(1)^2$ of the $U(1)^3$ isometry of the
six-dimensional space. Although the full $U(1)^3$ is preserved for
$\delta=-1/2$, the form field would blow up at the degeneracies of
$X$, $Y$ and $Z$, which would lead to a singular surface in the
ten-dimensional geometry. In order for the singularity to be confined
to $X=0$, so that we have a reasonable gravity description near the UV
region of the dual field theory, we require that $\delta\ge 0$.

We find there exist the following $(2,1)$-forms: 
\bea
\Phi_1 &=& f\Big( \fft{Y Z}{X}\, e^{\im\, 2\nu}\Big) \,
\fft{e^{\im\, \nu}}{X}\, \bar\epsilon^1\wedge \epsilon^2
   \wedge\epsilon^3\,,\nn\\
\Phi_2 &=& f\Big( \fft{X Z}{Y}\, e^{\im\, 2\nu}\Big) \,
\fft{e^{\im\, \nu}}{Y}\, \epsilon^1\wedge\bar\epsilon^2
   \wedge\epsilon^3\,,\nn\\ \Phi_3 &=& f\Big( \fft{X Y}{Z}\,
e^{\im\, 2\nu}\Big) \,  \fft{e^{\im\, \nu}}{Z}\, \epsilon^1\wedge\epsilon^2
   \wedge\bar\epsilon^3\,,\\
\Phi_4 &=& a_1 A\, \epsilon^1\wedge (\bar \epsilon^2\wedge \epsilon^2 -
\bar\epsilon^3\wedge \epsilon^3) +
a_2 B\, \epsilon^2\wedge (\bar\epsilon^3\wedge\epsilon^3 -
\bar\epsilon^1\wedge \epsilon^1)\nn\\ && +\,
a_3 C\, \epsilon^3\wedge (\bar\epsilon^1\wedge\epsilon^1 -
\bar\epsilon^2\wedge\epsilon^2)\,,\nn\\
\Phi_5 &=& b_1 A x\, \epsilon^1\wedge (\bar \epsilon^2\wedge \epsilon^2 -
\bar\epsilon^3\wedge \epsilon^3) +
b_2 B y\, \epsilon^2\wedge (\bar\epsilon^3\wedge\epsilon^3 -
\bar\epsilon^1\wedge \epsilon^1) \nn\\ && +\,
b_3 C z\, \epsilon^3\wedge (\bar\epsilon^1\wedge\epsilon^1 -
\bar\epsilon^2\wedge\epsilon^2)\,,\nn\label{Lpqrforms}
\eea
where
\bea
A^{-1}&=&(y-z)^2\sqrt{(y-x)(z-x)X}\,,\qquad
B^{-1}=(x-z)^2\sqrt{(x-y)(z-y)Y}\,,\nn\\
C^{-1}&=&(x-y)^2\sqrt{(x-z)(y-z)Z}\,,
\eea
and $a_i$ and $b_i$ are constants which satisfy the conditions $a_1 +
a_2 + a_3=0$ and $b_1+ b_2 + b_3=0$.  Notice that the first three
forms in (\ref{Lpqrforms}) are related to each other by interchanging
the $x$, $y$ and $z$ coordinates, while the last two forms remain
invariant. This reflects the fact that the $x$, $y$ and $z$
coordinates appear in a completely symmetric manner in the metric of
the resolved cone over $L^{abc}$. $\Phi_1$ has a singularity that is
confined to small distance, as do $\Phi_4$ and $\Phi_5$ if one
performs the rescaling $y\rightarrow \alpha y$, $z\rightarrow \alpha
z$ and then takes the limit $\alpha\rightarrow 0$. $\Phi_4$ and
$\Phi_5$ have nontrivial flux, while $\Phi_1$ does not.

In the cohomogeneity-two limit, the resolved $L^{abc}$ cones reduce to
the resolved $Y^{pq}$ cones. In this limit, $\Phi_4$ and $\Phi_5$
reduce to the corresponding forms given in (\ref{Ypqforms}), while the
first three forms generalize those in (\ref{Ypqforms}) to include an
arbitrary function $f$. In particular, taking $f=1$ reproduces the
$\Phi_1$ and $\Phi_2$ in (\ref{Ypqforms}), whilst taking $f$ to be the
inverse of its argument reproduces $\Phi_3$.

\section{D3-branes and the AdS/CFT correspondence}

A supersymmetric D3-brane solution of the type IIB theory
with six-dimensional Calabi-Yau transverse space is given by
\bea
ds&=&H^{-\ft12} (-dt^2 + dx_1^2 + dx_2^2 + dx_3^2) + H^{\ft12} ds_6^2
\,,\nn\\
F_5&=& G_\5 + * G_\5\,,\qquad
G_\5=dt\wedge dx_1\wedge dx_2\wedge dx_3 \wedge dH^{-1}\,,\nn\\
F_\3 &=& F_\3^{\rm RR} + {\rm i}\, F_\3^{\rm NS} = m\, \omega_\3\,, \label{D3soln}
\eea
with
\be
\square_6 H= m^2 |\omega_\3|^2\,.
\label{Heqn}
\ee
Here the $\square_6$ is a Laplacian of the Calabi-Yau metric $ds_6^2$
and $\omega_\3$ is a harmonic $(2,1)$-form in $ds_6^2$.  We shall
refer to this as a modified D3-brane solution, owing to the inclusion
of the additional 3-form. If this 3-form carries nontrivial flux, then
it corresponds to fractional a D3-brane.

We shall take the six-dimensional metric $ds_6^2$ of the transverse
space to be the resolved cone over $L^{abc}$. We first consider the
case of vanishing $m$.  It was shown in \cite{separability1,separability2,separability3} that the
Klein-Gordon equation for the general AdS-Kerr-NUT solutions
constructed in \cite{nutbh2} is separable.  Since our metrics arise as
the Euclideanization of the supersymmetric limit of AdS-Kerr-NUT
solutions, the corresponding equation for $H$ is hence also separable.
To see this, we consider a real superposition of the ansatz
\be
H=H_1(x)\, H_2(y)\, H_3(z)\, e^{2{\rm i} (a_0 \psi -a_1 \phi +
a_2\, \tau)}\,.
\ee
In general, this ansatz breaks the  $U(1)^3$ global symmetry.

The Laplace equation is then given by
\bea
0&=&\fft{1}{(y-x)(z-x)}\Big(\fft{(X\, H_1')'}{H_1} - 
\fft{(a_0 + a_1 x+ a_2 x^2)^2}{X}\Big)\nn\\
&&+\fft{1}{(x-y)(z-y)}\Big(\fft{(Y\, H_2')'}{H_2} -
\fft{(a_0 + a_1 y+ a_2 y^2)^2}{Y}\Big)\nn\\
&&+\fft{1}{(x-z)(y-z)}\Big(\fft{(Z\, H_3')'}{H_3} -
\fft{(a_0 + a_1 z+ a_2 z^2)^2}{Z}\Big)\,,
\eea
where a prime denotes a derivative with respect to the separated
variable associated with the function $H_i$.  This equation can be
expressed as three separate equations in $x$, $y$ and $z$:
\bea
(X\, H_1')' -\Big(\fft{(a_0 + a_1 x+ a_2 x^2)^2}{X} +
          b_0 + b_1 x\Big) H_1 &=& 0\,,\nn\\
(Y\, H_2')' -\Big(\fft{(a_0 + a_1 y+ a_2 y^2)^2}{Y} +
          b_0 + b_1 y\Big) H_2 &=& 0\,,\nn\\
(Z\, H_3')' -\Big(\fft{(a_0 + a_1 z+ a_2 z^2)^2}{Z} +
          b_0 + b_1 z\Big) H_3 &=& 0\,,\label{separateeqns}
\eea
where $b_0$ and $b_1$ are separation constants.  These equations do
not have explicit closed-form solutions for general $a_i$ and $b_i$.
We shall consider the simplest solution obtained by setting all of the
$a_i$ and $b_i$ to zero and letting $H$ depend on $x$ only.  The
solution is given by
\be
H=c_0 - \fft{c_1\,\log(x-x_1)}{(x_1-x_2)(x_1-x_3)} +
        \fft{c_1\,\log(x-x_2)}{(x_2-x_1)(x_2-x_3)} +
        \fft{c_1\,\log(x-x_3)}{(x_3-x_1)(x_3-x_2)}\,. \label{divergence1}
\ee
where $x_1,x_2$ and $x_3$ are the three roots of $X$, satisfying
\be
x_1 + x_2 + x_3=\alpha +\beta\,,\qquad
x_1 x_2 + x_1 x_3 + x_2 x_3 = \alpha\,\beta\,,\qquad
x_1 x_2 x_3=2 M\,.
\ee
Consider the radial coordinate $x$ with $x_1\le x\le\infty$. Then the
function $H$ has a logarithmic divergence at $x_1$.

We now consider solutions for which the 3-form $\omega_\3$ is turned
on.  A simple solution can be obtained by rescaling $y\rightarrow
\alpha y$, $z\rightarrow \beta z$ and then taking the limit
$\alpha=\beta=0$. The general construction of \cite{bergery,pagepope}
is recovered for the case of this class of resolved $L^{abc}$ cones
\cite{sfetsos}. We can then take $\omega_\3$ to be the harmonic
$(2,1)$-form $\Psi_1$ given by (\ref{psi1}). Then, for a certain
choice of integration constants, the resulting $H$ is given by
\be
H=\frac{x}{18M(x^3-2M)}\,, \label{divergence2}
\ee 
which diverges at $x^3=2M$. 

The divergence of the $H$ function in both (\ref{divergence1}) and
(\ref{divergence2}) corresponds to a naked singularity in the
short-distance region of the geometry. This singularity of the
D3-brane solution arises even in the case of the resolved cone over
$Y^{2,1}$, which itself is completely regular \cite{oota}. This singularity 
is due to the fact that the D3-branes have been smeared over the blown-up 
4-cycle. A shell of uniformly distributed branes tends to be singular at its 
surface. For the case of the resolved conifold, in which there is a blown-up 
2-cycle, a completely regular solution has been found for which the 
D3-branes are stacked at a single point \cite{murugan}. This involves 
solving the equations (\ref{separateeqns}) for the case of $T^{1,1}$ for 
which there is a delta function source. The solution is expressed as an 
expansion in terms of the angular harmonics. It would be interesting to 
explore than analogous construction for the resolved $L^{abc}$ cones. All 
of these other examples, with the sole exception of $Y^{2,1}$, will still have 
orbifold singularities.

Another possible way in which regular solutions can be obtained is to blow 
up a 3-cycle instead of a 4-cycle. Then an appropriate 3-form would prevent 
the 3-cycle from collapsing, as in the case of the deformed conifold \cite{ks}. 
As already discussed in the previous section, while there exists an 
obstruction to complex deformations of $Y^{pq}$ cones there are other 
subsets of the $L^{abc}$ cones which do allow for complex structure 
deformations \cite{altmann1,altmann2,LpqrCFT2}.  However, the explicit 
metrics for these deformed $L^{abc}$ cones are not known.

Although the solution describing D3-branes on a resolved $L^{abc}$
cone becomes singular at short distance, we can still use this
background at large distance to study various flows of the quiver
gauge theory in the region of the UV conformal fixed point. At large
$x$, (\ref{Heqn}) becomes
\be
\frac{4}{x^2} \partial_x\Big( X \partial_x H\Big) =m^2 |\omega_\3|^2\,,
\ee
where $X$ is given by (\ref{uvwXYZ}).  Note that this equation applies
for arbitrary $\alpha$ and $\beta$, since for large $x$ we can
consistently neglect the dependence of $H$ on the non-azimuthal
``angular" coordinates $y$ and $z$. Again considering the case of the
self-dual harmonic $(2,1)$-form $\Psi_1$ given by (\ref{psi1}), the
resulting asymptotic expansion of $H$ is
\be
H=\fft{Q}{x^2}\Big( 1+\fft{c_2}{x}+\fft{c_4}{x^2} +\fft{c_6}{x^3}+
\cdots\Big)\,,
\label{Hexpansion}
\ee
where 
\bea
c_2 &=& \fft23 (\alpha+\beta)\,,\nn\\ c_4 &=& \fft12
(\alpha^2+\alpha\beta+\beta^2)\,,\nn\\ c_6 &=& \fft{1}{30}
\Big( \fft{m^2}{Q}+12(\alpha^2+\beta^2)(\alpha+\beta)+2M\Big)\,.
\eea
We have set an additive constant to zero so that the geometry is
asymptotically AdS$_5\times L^{abc}$. This can be seen from the
leading $x^{-2}\sim r^{-4}$ term in $H$ (since $x$ has dimension two,
we can take $x\sim r^2$ for large $x$). The transformation properties and dimensions of the
operators being turned on in the dual field theory can be read off
from the linearized form of the supergravity solution (\ref{D3soln}). The metric perturbations due to $H$ have the same form as those within the metric $ds_6^2$ itself. Therefore, from the asymptotic expansion of $H$ given in (\ref{Hexpansion}), we can read off that there are
scalar operators of dimension two, four and six with
expectation values that go as  $c_2$, $c_4$ and $c_6$, respectively. This is consistent with the perturbations of the 2-form and 4-form potentials. We shall now
discuss the gauge theory interpretation of the blown-up 2-cycles, as
well as the 3-form, in more detail.

\bigskip
\noindent\underline{Blown-up 2-cycle}
\bigskip

First, we consider the case with vanishing $M$, for which the
six-dimensional space is the $L^{abc}$ analog of the resolved
conifold, in the sense that there is a blown-up 2-cycle. The volume of
the 2-cycle is characterized by the parameters $\alpha$ and
$\beta$. This is a global deformation, in that it changes the position
of the branes at infinity \cite{leo}.

The parameters $\alpha$ and $\beta$ specify the expectation values
of dimension $n$ non-mesonic scalar operators in the dual gauge
theory. For the case $\beta=-\alpha$, $c_2$ and $c_6$ vanish, while
$c_4$ can only vanish for $\alpha=\beta=0$. To identify the specific
dimension-two operator whose expectation value goes as $c_2$, it is helpful
to consider the description of the resolved cone over $L^{abc}$ in
terms of four complex numbers $z_i$ which satisfy the constraint
\be
\sum_{i=1}^4 Q_i\, |z_i|^2=t\,, \label{complex}
\ee
where one then takes the quotient by a $U(1)$ action
\cite{martelli}. The parameter $t$ is the area of the blown-up $CP^1$
and corresponds to the coefficient of the Fayet-Iliopoulos term in the
Lagrangian of the field theory. The $z_i$ correspond to the lowest
components of chiral superfields. This can be described as a gauged
linear sigma model with a $U(1)$ gauge group and 4 fields with charges
$Q_i$.  Then the above constraint corresponds to setting the D-terms
of the gauged linear sigma model to zero to give the vacuum. For the
$L^{abc}$ spaces, the $Q_i$ are given by $Q_i =(a,-c,b,-d)$ where
$d=a+b-c$ \cite{LpqrCFT1}. The requirement $\sum_{i=1}^4 Q_i=0$
guarantees that the 1-loop $\beta$-function vanishes, so that the
sigma model is Calabi-Yau.

Since $t$ acts as a natural order-parameter in the gauge theory, from
(\ref{complex}) it is reasonable to suppose that blowing up the
2-cycle corresponds to giving an expectation value that goes as $c_2$ to the
dimension-two scalar operator\footnote{We thank Amihay Hanany and Igor
Klebanov for correspondence on this point.}
\be
{\cal K}=a\, A_{\alpha} {\bar A}^{{\dot\alpha}}-c\,
B_{{\dot\alpha}} {\bar B}^{{\dot\alpha}}+ b\, C_{\alpha}
{\bar C}^{\alpha}-d\, D_{{\dot\alpha}} {\bar D}^{{\dot\alpha}}\,.
\ee
This operator lies within the $U(1)$ baryonic current multiplet. Since
this conserved current has no anomalous dimension, the dimension of
${\cal K}$ is protected. ${\cal K}$ reduces to the operator discussed
in \cite{leo} for the case of a resolved cone over $T^{11}/\Z_2$, for
which $a=b=c=d=1$.

\bigskip
\noindent\underline{Blown-up 4-cycle}
\bigskip

For nonvanishing $M$ in the function $X$, one generically blows up a
4-cycle. Unlike the case of a blown-up 2-cycle, this is a local
deformation since it does not change the position of the branes at
infinity \cite{leo}. In the limit of vanishing $\alpha$ and $\beta$,
one recovers the general construction obtained in
\cite{bergery,pagepope} that has been recently discussed in
\cite{pal,sfetsos,leo}. Also note that $c_6$ vanishes for the
appropriate values of $M$, $\alpha$ and $\beta$.

It has been shown that the number of formal Fayet-Iliopoulos
parameters can be matched with the possible deformations, which is
suggestive that the dimension-six operator that is turned on is
associated with the gauge groups in the quiver. Although the specific
operator has not been identified, it has been proposed that they are
of the schematic form \cite{leo}
\be
{\cal O}_i=\sum_g c_{i,g} {\cal W}_g {\bar {\cal W}}_g\,,  \label{O1}
\ee
where the gauge groups in the quiver have been summed over, ${\cal
W}_g$ is an operator associated with the field strength for the gauge
group $g$, and $c_{i,g}$ are constants. The dimension-six operator
might also have contributions from the bifundamental fields of the
form
\be
a_1 A_{\alpha} {\bar A}^{\alpha} B_{{\dot\alpha}} {\bar
B}^{{\dot\alpha}} C_{\beta} {\bar C}^{\beta} +a_2 A_{\alpha} 
{\bar A}^{\alpha} B_{{\dot\alpha}} {\bar B}^{{\dot\alpha}}
D_{\dot\beta} {\bar D}^{\dot\beta} +a_3 A_{\alpha} {\bar A}^{\alpha}
C_{\beta} {\bar C}^{\beta} D_{\dot\beta} {\bar D}^{\dot\beta}
+a_4 B_{\dot\alpha} {\bar B}^{{\dot\alpha}} C_{\beta}
{\bar C}^{\beta} D_{\dot\beta} {\bar D}^{\dot\beta}, \label{O2}
\ee
where the $a_i$ are constants. It is proposed that a particular
combination of all of these terms in (\ref{O1}) and (\ref{O2})
correspond to the blown-up 4-cycle\footnote{We thank Sergio Benvenuti
for correspondence on this point.}. One possibility is that the
contributions from the bifundamental fields in (\ref{O2}) are present
only when $\alpha$ and $\beta$ are nonvanishing.

\bigskip
\noindent\underline{Turning on the 3-form}
\bigskip

Turning on a 3-form results in the ranks of some of the gauge groups
of the dual quiver gauge theory being reduced with decreasing energy
scale. For the case in which the 3-form has nontrivial flux, the
theory undergoes a Seiberg duality cascade \cite{frac1,frac2,ks}. On
the other hand, the 3-form $\Psi_1$ given by (\ref{psi1}) does not
have nontrivial flux. For a case such as this, it has been proposed
that the reduction in ranks of gauge groups is due to Higgsing
\cite{aharony}. In particular, from (\ref{Hexpansion}), we see that
the parameter $m$ associated with the 3-form also contributes to the
expectation value $c_6$ of a dimension-six scalar operator. An
additional effect of this 3-form is that the $U(1)$ R-symmetry is
broken.  The theory still preserves ${\cal N}=1$ supersymmetry.

\section{Eight-dimensional resolved Calabi-Yau cones}

\subsection{Cohomogeneity-two metrics}

We now turn to eight-dimensional Calabi-Yau spaces, which can be
used to construct M2-brane solutions of eleven-dimensional
supergravity. Before considering the general cohomogeneity-four
resolved cones over $L^{pqrs}$, we shall first look at the
cohomogeneity-two metrics, which can be built over an $S^2\times S^2$
base space. These metrics are given by \cite{nutbh1,nutbh2}
\bea
ds_8^2 &=& \ft14 u^2 dx^2 + \ft14 v^2 dy^2 +
\fft{1}{u^2} \Big[d\tau + \fft{y}{3\alpha}
(\sigma_3 + \td \sigma_3)\Big]^2\nn\\
&&+
\fft{1}{v^2}\Big[d\tau - \fft{x}{3\alpha}(\sigma_\3 +\td \sigma_3)\Big]^2
+c^2 (\sigma_1^2 + \sigma_2^2 +\td \sigma_1^2 +
\td \sigma_2^2)\nn\\
u^2&=&\fft{x+y}{X}\,,\qquad 
v^2=\fft{x+y}{Y}\,,\qquad c^2=\fft{xy}{6\alpha}\,,\nn\\
X&=&x(x+\alpha) - \fft{2\mu}{x^2}\,,\quad
Y=y(\alpha-y) + \fft{2\nu}{y^2}\,,\label{coho2d8}
\eea
Completely regular examples were discussed in \cite{lupores}.

\newpage
\noindent\underline{Calabi-Yau structure}
\bigskip

We can define the vielbein basis
\bea
e^1&=&\ft12 u\, dx\,,\qquad  e^2=-\fft{1}{u}(d\tau +\fft{y}{3\alpha}
(\sigma_3 + \td\sigma_3))\,,\qquad 
e^3=\ft12 v\, dy\,,\\
e^4&=&\fft{1}{v}(d\tau -\fft{x}{3\alpha}
(\sigma_3 + \td\sigma_3))\,,\quad
e^5=c\,\sigma_1\,,\quad e^6=c\,\sigma_2\,,\qquad
e^7=c\,\td\sigma_1\,,\quad e^8=c\,\td\sigma_2\,,\nn
\eea
and then the complex vielbein
\bea
\epsilon_1=e^1 + {\rm i}\, e^2\,,\qquad
\epsilon_2=e^3 + {\rm i}\, e^4\,,\qquad
\epsilon_3=e^5 + {\rm i}\, e^6\,,\qquad
\epsilon_4=e^7 + {\rm i}\, e^8\,.
\eea
The K\"ahler 2-form and holomorphic $(4,0)$-form are given by
\be
J=\ft{\rm i}{2} \epsilon^i\wedge \bar \epsilon^i\,,
\ee
and
\be
G_\4=e^{-4{\rm i}\tau} \epsilon^1\wedge\epsilon^2\wedge
\epsilon^3\wedge \epsilon^4\,.
\ee

\bigskip
\noindent\underline{Harmonic $(2,2)$-forms}
\bigskip

We find four self-dual $(2,2)$-forms; they are given by
\crampest{
\bea
\Phi_1&=&\fft{(\bar\epsilon_1\wedge \epsilon_1 +
\bar\epsilon_2 \wedge \epsilon_2)\wedge(\bar \epsilon_3\wedge
\epsilon_3 + \bar \epsilon_4\wedge\epsilon_4) -
2(\bar\epsilon_1\wedge\epsilon_1\wedge\bar\epsilon_2\wedge\epsilon_2
+\bar\epsilon_3\wedge\epsilon_3\wedge\bar\epsilon_4\wedge\epsilon_4)
}{x^3y^3}\nn\\
\Phi_2&=&\fft{(\bar \epsilon_1\wedge\epsilon_1 -
\bar\epsilon_2\wedge\epsilon_2)\wedge (\bar \epsilon_3\wedge
\epsilon_3 - \bar\epsilon_4\wedge\epsilon_4)}{xy(x+y)^2}\,,\nn\\
\Phi_3&=& \fft{e^{-4{\rm i}\tau}\,
(\bar\epsilon_1\wedge\bar\epsilon_2
\wedge\epsilon_3\wedge\epsilon_4 +\epsilon_1\wedge\epsilon_2
\wedge\bar\epsilon_3\wedge\bar\epsilon_4)}{x^2y^2\,XY}\,,\nn\\
\Phi_4&=&\fft{(\bar\epsilon_1\wedge\epsilon_2 - \epsilon_1\wedge
\bar\epsilon_2)\wedge (\bar \epsilon_3 \wedge\epsilon_3 +
\bar\epsilon_4\wedge\epsilon_4)}{xy\sqrt{XY}}\,.
\label{22forms}
\eea
}
Notice that $\Phi_1$ and $\Phi_2$ are square integrable,
in that they are well behaved at both small and large asymptotic distance. 
For the cases in which the eight-dimensional Calabi-Yau spaces are 
regular \cite{lupores}, these harmonic forms can be used to construct 
completely non-singular M2-brane solutions to eleven-dimensional supergravity.

\subsection{Cohomogeneity-four metrics on resolved cones over $L^{pqrs}$}

We now turn to the general cohomogeneity-four metrics on resolved
Calabi-Yau cones over the seven-dimensional Einstein-Sasaki spaces
$L^{pqrs}$, which can be written as \cite{nutbh2}
\crampest{
\bea
ds_8^2 &=& \ft14 (u_1^2 \, dx_1^2 + u_2^2\,  dx_2^2 + u_3^2 \, dx_3^2 
                   + u_4^2\, dx_4^2) \nn\\
&&+\fft1{u_1^2}\, [d\tau + (x_2+x_3+x_4) d\phi + 
      (x_2 x_3 + x_2 x_4 + x_3 x_4) d\psi + x_2 x_3 x_4 d\chi]^2 \nn\\
&&+\fft1{u_2^2}\, [d\tau + (x_1+x_3+x_4) d\phi +
      (x_1 x_3 + x_1 x_4 + x_3 x_4) d\psi + x_1 x_3 x_4 d\chi]^2 \nn\\
&&+\fft1{u_3^2}\, [d\tau + (x_1+x_2+x_4) d\phi +
      (x_1 x_2 + x_1 x_4 + x_2 x_4) d\psi + x_1 x_2 x_4 d\chi]^2 \nn\\
&&+\fft1{u_4^2}\, [d\tau + (x_1+x_2+x_3) d\phi +
      (x_1 x_2 + x_1 x_3 + x_2 x_3) d\psi + x_1 x_2 x_3 d\chi]^2\,,
\eea}
where
\bea
u_1^2 &=& \fft{(x_2-x_1)(x_3-x_1)(x_4-x_1)}{X_1}\,,\qquad
u_2^2 = \fft{(x_1-x_2)(x_3-x_2)(x_4-x_2)}{X_2}\,,\nn\\
u_3^2 &=& \fft{(x_1-x_3)(x_2-x_3)(x_4-x_3)}{X_3}\,,\qquad
u_4^2 = \fft{(x_1-x_4)(x_2-x_4)(x_3-x_4)}{X_4}\,,\nn\\
X_1&=& x_1(a-x_1)(b-x_1)(c-x_1) -2 M_1\,,\nn\\
X_2 &=&  x_2(a-x_2)(b-x_2)(c-x_2) -2 M_2\,,\nn\\
X_3&=& x_3(a-x_3)(b-x_3)(c-x_3) -2 M_3\,,\nn\\
X_4 &=&  x_4(a-x_4)(b-x_4)(c-x_4) -2 M_4\,.
\eea

\bigskip
\noindent\underline{Calabi-Yau structure}
\bigskip

    We shall choose the vielbein basis
\bea
e^1&=& \ft12 u_1 \, dx_1\,,\quad u_3=\ft12 u_2\,dx_2\,,\quad
  e^5= \ft12 u_3\, dx_3\,,\quad e^7= \ft12 u_4\, dx_4\,,\nn\\
e^2 &=& \fft1{u_1}\, [d\tau + (x_2+x_3+x_4) d\phi +
      (x_2 x_3 + x_2 x_4 + x_3 x_4) d\psi + x_2 x_3 x_4 d\chi]\,,\nn\\
e^4&=& \fft1{u_2}\, [d\tau + (x_1+x_3+x_4) d\phi +
      (x_1 x_3 + x_1 x_4 + x_3 x_4) d\psi + x_1 x_3 x_4 d\chi]\,,\nn\\
e^6 &=&\fft1{u_3}\,  [d\tau + (x_1+x_2+x_4) d\phi +
      (x_1 x_2 + x_1 x_4 + x_2 x_4) d\psi + x_1 x_2 x_4 d\chi]\,,\nn\\
e^8 &=& \fft1{u_4}\, [d\tau + (x_1+x_2+x_3) d\phi +
      (x_1 x_2 + x_1 x_3 + x_2 x_3) d\psi + x_1 x_2 x_3 d\chi]\,.
\eea
 The holomorphic vielbein are then given by
\be
\ep^1= e^1 + \im\, e^2\,,\quad \ep^2= e^3+\im\, e^4\,,\quad
  \ep^3 = e^5+\im\, e^6\,,\quad \ep^4= e^7+\im\, e^8\,.
\ee
Defining
\bea
J&=& \fft{\im}{2}\, (\bar\ep^1\wedge \ep^1 + \bar\ep^2\wedge\ep^2 + 
         \bar\ep^3\wedge\ep^3 + \bar\ep^4\wedge \ep^4)\,,\nn\\
\Omega &=& e^{\im\, \nu}\, \ep^1\wedge\ep^2\wedge\ep^3 \wedge\ep^4\,,
\eea
where
\be
\nu= 4\tau + 3(a+b+c)\, \phi + 2 (ab + bc + ca)\, \psi + abc\, \chi\,,
\ee
it is straightforward to verify that
\be
dJ=0\,,\qquad d\Omega=0\,,
\ee
and hence that the metric is indeed Ricci-flat K\"ahler, with $J$
being the K\"ahler form and $\Omega$ the holomorphic $(4,0)$-form.

\bigskip
\noindent\underline{Harmonic $(3,1)$-forms}
\bigskip

   We find that harmonic $(3,1)$-forms can be constructed as follows.
First, it can be verified that
\be
G_{(3,1)}= \fft1{X_1}\, e^{\im\, \nu}\, \bar\ep^1\wedge\ep^2\wedge\ep^3
\wedge\ep^4
\ee
is closed, and hence harmonic.  Next, we define the function 
\be
\gamma= \sqrt{\fft{X_2 X_3 X_4}{X_1}}\, e^{\im\, \nu}\,,
\ee
which can be shown to satisfy the relation
\bea
d\gamma &=& \fft{u_1\, e^{\im\, \nu}}{u_2 u_3 u_4\, 
  (x_1-x_2)(x_1-x_3)(x_1-x_4)}\, \Big( u_1 \,(x_2-x_3)(x_2-x_4)(x_4-x_3) 
        X_1'\, \bar\ep^1 \nn\\
&& - u_2 \, (x_3-x_1)(x_3-x_4)(x_4-x_1)\, X_2'\, \ep^2 +
u_3\, (x_1-x_2)(x_4-x_1)(x_4-x_2)\, X_3'\, \ep^3\nn\\
&& + u_4\, (x_1-x_2)(x_3-x_1)(x_2-x_3)\, X_4'\, \ep^4\Big)\,,
\eea
where $X_i'$ denotes the derivative of $X_i$ with respect to its
argument $x_i$.  It therefore follows that $d\gamma\wedge G_{(3,1)}=0$,
and so
\be
\Phi_{(3,1)} = f(\gamma)\, G_{(3,1)}
\ee
is a harmonic $(3,1)$-form for any function $f$.  In particular, we have
a family of harmonic $(3,1)$-forms given by
\be
\Psi_{(3,1)} = \fft{X_2^\delta\, X_3^\delta\, X_4^\delta}{X_1^{\delta+1}}\, 
 e^{(2\delta+1)\, \im\, \nu}\,  \bar\ep^1\wedge\ep^2\wedge\ep^3
\wedge\ep^4
\ee
for any constant $\delta$. For nonzero $\delta$, these forms preserve 
only a $U(1)^3$ subgroup of the $U(1)^4$ isometry of the eight-dimensional 
space. Note that $\Psi_{(3,1)}$ has a singularity only at short distance 
if $\delta\ge 0$, where we have taken $x_1$ to be the radial 
direction. Additional harmonic $(3,1)$-forms can be constructed by 
permuting the $x_i$ directions, but these forms have singularities 
for all $x_1$.
They are analogous to the $(2,1)$-forms $\Phi_1$, 
$\Phi_2$ and $\Phi_3$ in (\ref{Lpqrforms}) for a six-dimensional space,
and they do not support nontrivial flux.

\bigskip
\noindent\underline{Harmonic $(2,2)$-forms}
\bigskip

  We can also construct harmonic $(2,2)$-forms as follows.  We define
$(2,2)$-forms 
\bea
G_{(2,2)} &=& f\, (\bar\ep^1\wedge \ep^1 \wedge \bar\ep^2\wedge\ep^2 
                 + \bar\ep^3\wedge \ep^3 \wedge \bar\ep^4\wedge\ep^4)\nn\\
 &&+ g\, (\bar\ep^1\wedge \ep^1 \wedge \bar\ep^3\wedge\ep^3 
                 + \bar\ep^2\wedge \ep^2 \wedge \bar\ep^4\wedge\ep^4)\nn\\
&& + h\,  (\bar\ep^1\wedge \ep^1 \wedge \bar\ep^4\wedge\ep^4
                 + \bar\ep^2\wedge \ep^2 \wedge \bar\ep^3\wedge\ep^3)\,,
\label{G4ans}
\eea
where $f$, $g$ and $h$ are functions of $(x_1,x_2,x_3,x_4)$.  Imposing
the closure of $G_{(2,2)}$ leads to three independent solutions for
$f$, $g$ and $h$, namely
\be
f=g=h=1\,,\label{4sol1}
\ee
\bea
f &=& \fft{1}{(x_1-x_2)^2 (x_1-x_3)(x_2-x_4)(x_3-x_4)^2}\,,\nn\\
g &=& \fft{x_1\, (2 x_4-x_2- x_3) + x_2\, (2 x_3-x_4) - x_3\, x_4}{
   (x_1-x_2)^2 (x_1-x_3)^2 (x_2-x_4)^2 (x_3-x_4)^2}\,,\nn\\
h &=& \fft{1}{(x_1-x_2)(x_1-x_3)^2 (x_2-x_4)^2 (x_3-x_4)}
\,,\label{4sol2}
\eea
and 
\bea
f&=& \fft1{(x_1-x_3)(x_2-x_3)^2(x_1-x_4)^2(x_2-x_4)} \,,\nn\\
g &=& \fft1{(x_1-x_3)^2 (x_2-x_3)(x_1-x_4)(x_2-x_4)^2}\,,\nn\\
h&=& \fft{x_1\, (x_3+x_4-2x_2) + x_2\, (x_3+x_4) - 2 x_3 \, x_4}{
        (x_1-x_3)^2 (x_2-x_3)^2 (x_1-x_4)^2 (x_2-x_4)^2}\,.\label{4sol3}
\eea

These forms are somewhat analogous to the $(2,1)$-forms $\Phi_4$ 
and $\Phi_5$ given in (\ref{Lpqrforms}) for a six-dimensional space. 
The first solution, (\ref{4sol1}), is just the harmonic $(2,2)$-form
$J\wedge J$.  It follows from (\ref{G4ans}) that $J\wedge G_{(2,2)}$
is proportional to $(f+g+h)$, and so $J\wedge G_{(2,2)}$ is non-zero
for (\ref{4sol1}).  However, each of the solutions (\ref{4sol2}) and
(\ref{4sol3}) satisfies $f+g+h=0$, and so these two harmonic
$(2,2)$-forms satisfy the supersymmetric condition
\be
J\wedge G_{(2,2)}=0\,.
\ee
Notice also that these harmonic $(2,2)$-forms are square integrable. 
These can be used to construct modified M2-brane solutions, which have 
only orbifold-type singularities. Note that none of these 
cohomogeneity-four Calabi-Yau spaces are completely regular \cite{lupores}.

\subsection{M2-brane solutions}

We can use these eight-dimensional spaces, and the harmonic 4-forms
which they support, to construct a modified M2-brane solution to
eleven-dimensional supergravity, given by
\bea
ds_{11}^2 &=& H^{-2/3} (-dt^2+dx_1^2+dx_2^2)+H^{1/3} ds_8^2\,,\nn\\
F_\4 &=& dt\wedge dx_1\wedge dx_2\wedge dH^{-1}+m\, L_\4\,,
\eea
where 
\be
\square H=-\fft{1}{48} m^2 L_\4^2\,,
\ee
and $L_\4$ is an (anti)self-dual harmonic 4-form on the
eight-dimensional space with the metric $ds_8^2$.

      Let us first consider the case with $m=0$, for which the Laplace
equation on the Calabi-Yau metric is separable.  The solution for
general dimensionality is presented in the appendix B.  Here we just
give a solution for the eight-dimensional case that depends only on
the radial variable $x_1$; it is given by
\be
H= \int^{x_1} \fft{3Q}{X(x_1')} dx_1'\,.
\ee
Thus in the asymptotic region at large $x_1$, the function $H$ has
the behavior
\be
H=\frac{Q}{x_1^3}\,\Big( 1 + \fft{c_2}{x_1}
+ \cdots\Big)\,,\qquad {\rm where}\ \ c_2=\fft34 (\alpha+\beta+\gamma)\,.
\ee
We have taken an arbitrary additive constant to zero, so that the
geometry is asymptotically AdS$_4\times L^{pqrs}$. Since $x_1$ has
dimension two, we see that there is a non-mesonic dimension-two scalar
operator being turned on with expectation value $c_2$.

It is especially interesting to construct M2-brane solutions using one
of the square-integrable harmonic $(2,2)$-forms that we found previously, 
since this guarantees that with the appropriate
integration constants the only singularities are of
orbifold type. This is because the 4-form prevents the blown-up
4-cycle from collapsing. Moreover, examples of regular 
eight-dimensional Calabi-Yau spaces that have been discussed in 
\cite{lupores} can be used to construct completely non-singular 
M2-brane solutions.
The resulting geometry smoothly interpolates
between AdS$_4\times L^{pqrs}$ asymptotically, and a direct product of
Minkowski$_3$ and a compact space at short distance.  Many examples of
cohomogeneity-one solutions of this type were constructed in
\cite{trans,ricciflat,hyper}. Although not much is known even about
the UV conformal fixed point of the dual three-dimensional ${\cal
N}=2$ super Yang-Mills field theory, based on the geometrical
properties of the supergravity background it flows to a confining
phase in the IR region.

\section{Harmonic forms on higher-dimensional resolved cones}

   In this section, we extend some of the constructions of harmonic
middle-dimension forms to the case of higher-dimensional metrics on
the resolutions of cones over Einstein-Sasaki spaces.  We take as our
starting point the local Ricci-flat K\"ahler metrics in dimension
$D=2n+4$ that were considered in \cite{lupores}:
\bea
d\td s^2 &=& \fft{x+y}{4 X} dx^2 + \fft{x+y}{4Y} dy^2 +
\fft{X}{x+y} \Big[d\tau + \fft{y}{\alpha}\sigma\Big]^2
+\fft{Y}{x+y}\Big[d\tau - \fft{x}{\alpha}\sigma\Big]^2 +
\fft{xy}{\alpha} d\Sigma_n^2\nn\\
\sigma&=& d\psi +A\,,\qquad
X=x(x+\alpha) - \fft{2\mu}{x^n}\,,\qquad
Y=y(\alpha-y) + \fft{2\nu}{y^n}\,,\label{coho2genmetric}
\eea
where $d\Sigma_n^2$ is a metric on a $2n$-dimensional
Einstein-K\"ahler space $Z$, satisfying $R_{ab}=2(n+1)\, g_{ab}$, with
K\"ahler form $J= \ft12 dA$.  (We have made some minor changes of
coordinates compared to the metric presented in \cite{lupores}.)  For
convenience, we shall set the constant $\alpha$ to unity.  This can
always be done, when $\alpha\ne0$, by means of coordinate scalings
together with an overall rescaling of the Ricci-flat metric.  The
special case $\alpha=0$ can be recovered via a limiting procedure.

   Next, we define the 2-forms
\be
\omega_x= \ft12 dx\wedge (d\tau + y\,\sigma)\,,\qquad
\omega_y= \ft12 dy\wedge (d\tau - x\,\sigma)\,,\qquad
\omega= x y\, J\,.
\ee
It can easily be verified that $\hat J\equiv \omega_x
-\omega_y+\omega$ is closed and, in fact, this is the K\"ahler form of
the Ricci-flat K\"ahler metric (\ref{coho2genmetric}).  In the case
that $n$ is even ($n=2m$), we find that the middle-degree form
\be
G_{(2m+2)}= \fft{1}{(x y)^{2m+1}}\, \Big[
   \omega_x\wedge\omega_y \wedge \omega^{m-1} +
  \fft1{m+1}\, (\omega_x-\omega_y)\wedge \omega^m -\fft1{m(m+1)}\,
      \omega^{m+1}\Big]
\ee
is closed.  Since it is also self-dual, it follows that it is a
harmonic form.  This generalises the harmonic $(2,2)$-form $\Phi_1$ in
eight dimensions given in (\ref{22forms}) and is somewhat analogous to 
the $(2,1)$-forms 
$\Phi_4$ and $\Phi_5$ given in (\ref{Lpqrforms}) for a six dimensions. 

   Further harmonic forms can be obtained if one
takes the Einstein-K\"ahler base metric $d\Sigma_n^2$ to be a product
of Einstein-K\"ahler metrics.  For example, if we choose it to be the
product of metrics on two copies of $\CP^m$ (recall that we are
considering the case where $n=2m$ is even), with K\"ahler forms $J_1$
and $J_2$ respectively (so $J=J_1+J_2$), then defining
\be
\omega_1= xy \, J_1\,,\qquad \omega_2 = x y \, J_2\,,
\ee
we find that
\be
\wtd G_{(2m+2)} = \fft{1}{(x+y)^2\, (xy)^m} \, (\omega_x+\omega_y)\wedge
  \sum_{p=0}^m (-1)^p \, \omega_1^{m-p}\wedge \omega_2^p
\ee
is closed and self-dual, and therefore it is harmonic.

\section{Conclusions}

We have investigated the K\"ahler moduli associated with blowing up a
2-cycle or 4-cycle on Calabi-Yau cones over the $L^{abc}$ spaces. This
yields a countably infinite number of backgrounds with ALE
singularities on which perturbative string dynamics is
well-defined. Although adding D3-branes induces a power-law type
singularity at short distance, one can still use the AdS/CFT
dictionary to relate the blown-up cycles to deformations of the dual
quiver gauge theory close to the UV conformal fixed point. In
particular, we identify the non-mesonic dimension-two real scalar
operator that acquires a vev, thereby generalizing the state/operator
correspondence for the resolved conifold over $T^{11}$
\cite{resolvedT11} and $T^{11}/\Z_2$ \cite{leo} to resolved cones over
the $L^{abc}$ spaces. On the other hand, blowing up a 4-cycle
corresponds to a dimension-six non-mesonic scalar operator getting a
vev.

The resolved cones over the cohomogeneity-two $L^{abc}$ spaces support
various harmonic $(2,1)$-forms, some of which depend nontrivially on
three non-azimuthal coordinate directions. These forms can be 
further generalized by a
multiplicative function, so long as the exterior derivative of this
function satisfies a certain orthogonality condition. In particular,
there are harmonic $(2,1)$-forms which depend on continuous
parameters.  3-forms carrying nontrivial flux correspond to fractional
D3-branes, while those which do not correspond to giving a vev to a
dimension-six operator.

For the D3-brane solutions constructed with resolved cones over $L^{abc}$, 
we have restricted ourselves to the case in which the D3-branes are smeared 
over the blown-up cycle. As we already mentioned, this yields to a power-law 
singularity at short distance. For solutions involving a 3-form field, one may 
be able to smooth out this singularity by a complex deformation of the 
Calabi-Yau space that results in a blown-up 3-cycle. Although it has been 
shown that there are obstructions to the existence of complex deformations 
of cones over $Y^{pq}$ spaces, there are other subsets of the $L^{abc}$ 
cones which do allow for complex structure deformations
\cite{altmann1,altmann2,LpqrCFT2}. It would be useful to construct the
explicit metrics describing these deformed $L^{abc}$ cones, as well as
the non-singular supergravity solutions that describe fractional
D3-branes on these spaces.

Alternatively, one can consider stacking the D3-branes at a single point. For 
the case of the resolved conifold, this has been shown to yield a completely 
regular solution \cite{murugan}. Perhaps there are analogous constructions  
with the resolved cones over $L^{abc}$. With the exceptions of 
$T^{1,1}$, $T^{1,1}/\Z_2$ and $Y^{2,1}$, the resolved $L^{abc}$ cones 
have orbifold singularities. Although these singularities will remain there 
when D3-branes are stacked at a single point, perturbative string dynamics 
is well-defined on such backgrounds.

One can also consider fibering a D3-brane worldvolume direction (which 
need not be compact) over a 
resolved $L^{abc}$ cone in such a way that the resulting geometry only has 
orbifold-type singularities. For the case of the resolved conifold, such a 
D3-brane solution has already been constructed and is completely regular, 
and it is also supersymmetric \cite{wrapped}. The corresponding D3-brane 
solutions for the resolved $L^{abc}$ cones are currently being investigated
\cite{work}.

We also discussed the geometry of higher-dimensional Calabi-Yau spaces
with blown-up cycles, as well as the various harmonic forms which live
on them. In particular, we have found that eight-dimensional resolved
cones over the $L^{pqrs}$ spaces support harmonic 4-forms that are
square integrable. They can be used to construct M2-brane solutions
of eleven-dimensional supergravity which have only orbifold-type
singularities. Unfortunately, not much is known about the dual
three-dimensional ${\cal N}=2$ gauge theories, other than that they
flow from a UV conformal fixed point to a confining phase in the IR
region.

Lastly, the type IIB supergravity backgrounds dual to certain marginal
deformations ($\beta$ deformations) of the conformal fixed point of
the $Y^{pq}$ and $L^{abc}$ quiver gauge theories were obtained in
\cite{lm,ahn}. The solution-generating method works for any gravity
solution with $U(1)\times U(1)$ global symmetry. It might be
interesting to see if these deformations can be applied to the gravity
solutions discussed in this paper, since they possess the necessary
global symmetry.

\section*{Acknowledgments}

We should like to thank Philip Argyres, Sergio Benvenuti, Aaron
Bergman, Davide Forcella, Amihay Hanany, Igor Klebanov, Jason Kumar, Louis Leblond,
Leopoldo Pando Zayas and Angel Uranga for helpful conversations and
correspondence. M.C. thanks the George P. \& Cynthia W. Mitchell
Institute for Fundamental Physics for hospitality during the course of
this work.

\appendix

\section{Complex structure and first-order equations}

   In this appendix, we construct Ricci-flat K\"ahler spaces in
dimension $D=2n+4$, built over an Einstein-K\"ahler base space of real
dimension $2n$ with metric $d\Sigma_n^2$.  We normalise this metric so
that it satisfies $R_{ij} = 2(n+1) g_{ij}$.  Its K\"ahler form will be
written as $J=\ft12 dA$.  We may also assume that it admits a
holomorphic $(n,0)$-form $\Omega$, satisfying (see, for example,
section 4 of \cite{lupopo})
\be
d\Omega= \im\, (n+1) A\wedge \Omega\,.\label{dOm}
\ee
The ansatz for the $(3n+4)$-dimensional Ricci-flat K\"ahler metrics will
be
\be
d\hat s^2 = u^2 dx^2 + v^2 dy^2 + a^2 (d\tau+f_1 \sigma)^2 +
   b^2 (d\tau+ f_2 \sigma)^2 + c^2 d\Sigma_n^2\,,\label{genmet}
\ee
where $a$, $b$, $c$, $u$, $v$, $f_1$ and $f_2$ are functions of $x$
and $y$, and
\be
\sigma = d\psi + A\,.\label{sigdef}
\ee

   We define the vielbein
\be
\hat e^1 = u dx\,,\qquad \hat e^2 = a(d\tau + f_1 \sigma)\,,\qquad
 \hat e^3= v dy\,,\qquad \hat e^4 = b (d\tau + f_2 \sigma)\,,\qquad
\hat e^i = c e^i\,,
\ee
where $e^i$ is a vielbein for the Einstein-K\"ahler base metric
$d\Sigma_n^2$.

   We make the ansatz
\be
\hat J= e^1\wedge e^2 + e^3\wedge e^4 + c^2 J
\ee
for the K\"ahler form.  It is then natural to define a complex
vielbein by
\be
\hat\ep^1 = \hat e^1 + \im\, \hat e^2\,,\qquad
\hat\ep^2 = \hat e^3 + \im\, \hat e^4\,,\qquad
\hat \ep^i = c\, \ep^i\,,
\ee
where $\ep^i$ is a complex vielbein for the base metric $d\Sigma_n^2$.
We also make the ansatz
\be
\hat\Omega= e^{\im\alpha \tau + \im\beta\psi}\, c^n\, \hat\ep^1
\wedge \hat\ep^2 \wedge\Omega\label{Omhat}
\ee
for the holomorphic $(n+2,0)$-form.  The conditions for $d\hat s^2$ to be
Ricci flat and K\"ahler are then given by
\be
d\hat J=0\,,\qquad d\hat\Omega=0\,.
\ee
One immediately finds that the constant
$\beta$ should be chosen to be
\be
\beta = n+1\,.
\ee
However, the constant $\alpha$ can be left arbitrary.

  We now obtain the first-order equations:
\bea
d\hat J=0:&&
    (bv)' - (au)\dot{} =0\,,\qquad
   (c^2)' - 2 a u f_1 =0\,,\qquad
   (c^2)\dot{} - 2bv f_2=0\,,\nn\\
d\hat\Omega=0:&&
   \alpha\, uv c^n - (av c^n)' - (bu c^n)\dot{} =0\,,\nn\\
&& \alpha bu c^n f_2 - (n+1) bu c^n + [ab c^n (f_1-f_2)]' =0\,,\nn\\
&& \alpha av c^n f_1 - (n+1) av c^n - [ab c^n (f_1-f_2)]\dot{} =0
\,.\label{genfo} \eea 
The constant $\alpha$ appearing in the
first-order equations (\ref{genfo}) is always trivial, in the sense
that it can be set to any chosen non-zero value without loss of
generality.  To see this, we perform the following rescaling of
coordinates and functions:
\bea
&&x\rightarrow \lambda \, x\,,\qquad
y\rightarrow \lambda\, y\,,\qquad
\tau \rightarrow \lambda \, \tau\,,\nn\\
&& c\rightarrow \lambda\, c\,,\qquad
f_1\rightarrow \lambda \, f_1
\qquad f_2\rightarrow \lambda \, f_2\,,
\eea
whilst leaving the functions $a$, $b$, $u$ and $v$ unscaled.  It can be
seen that the effect of these rescalings is to scale the metric $d\hat s^2$
in (\ref{genmet}) according to
\be
d\hat s^2 \rightarrow \lambda^2 \, d\hat s^2\,.
\ee
The rescalings have the effect of replacing $\alpha$ by
$\lambda\alpha$ in the first-order equations (\ref{genfo}), thus
giving
\bea
d\hat J=0:&&
    (bv)' - (au)\dot{} =0\,,\qquad
   (c^2)' - 2 a u f_1 =0\,,\qquad
   (c^2)\dot{} - 2bv f_2=0\,,\nn\\
d\hat\Omega=0:&&
    \lambda\alpha\, uv c^n - (av c^n)' - (bu c^n)\dot{} =0\,,\nn\\
&& \lambda\alpha\, bu c^n f_2 - (n+1) bu c^n + [ab c^n (f_1-f_2)]' =0\,,\nn\\
&& \lambda\alpha\, av c^n f_1 - (n+1) av c^n - [ab c^n (f_1-f_2)]\dot{} =0
\,.\label{genfo2}
\eea
Since a rescaling of a Ricci-flat metric by a non-zero constant leaves
it Ricci-flat, it follows that the constant $\lambda$ can be chosen at
will, and so no generality is lost by setting $\alpha$ to any desired
finite and non-zero value.

\section{Separability of Laplacian on Calabi-Yau metrics}

    We consider the Calabi-Yau metrics obtained in
\cite{nutbh1,nutbh2}.  The metric can be expressed as
\bea
ds^2&=&\sum_{\mu=1}^n \Big[\fft{U_\mu\, dx_\mu^2}{4X_\mu} +
\fft{X_\mu}{U_\mu}\,(\sum_{i=0}^{n-1} W_i d\phi_i)^2\Big]\,,\nn\\
X_\mu&=& x_\mu \prod_{i=1}^{n-1} (\alpha_i-x_\mu) - 2\ell_\mu\,,\qquad
U_\mu=\prod_{\nu=1}^{n}{}\!'\, (x_\nu-x_\mu)\,,
\eea
where $W_i$ is defined by
\be
\prod_{\mu=1}^n (1 + q x_\mu) \equiv \sum_{i=0}^{n-1} W_i\, q^{i+1}\,.
\ee
It turns out that the equation $\square H=0$ is separable in the
$x_{\mu}$ coordinates, where $\square$ is the Laplacian taken on the
above metric. (The separability for the more general
non-extremal Kerr-NUT-AdS metrics was shown explicitly in
\cite{separability1,separability2,separability3}.  Making the ansatz
\be
H=\Big(\prod_{\mu=1}^n H_\mu(x_\mu)\Big)\, \exp\Big(2{\rm i}\sum_{i=0}^{n-1}
(-1)^i a_i \phi_{n-1-i}\Big)\,,
\ee
for the harmonic function, we find that the $H_\mu(x_\mu)$ satisfy
\be
(X_\mu H_\mu')' - \Big(\fft{(\sum_{i=0}^{n-1} a_i\, x_\mu^i)^2}{X_\mu} +
\sum_{i=1}^{n-2} b_i x_\mu^i\Big)\,H_\mu=0\,,
\ee
where a prime on $H_\mu$ or $X_\mu$ denotes a derivative with respect to
its argument $x_\mu$.  The system thus has $2n-1$ independent
separation constants $a_0, a_1,\ldots a_{n-1}$ and $b_0, b_1, \ldots,
b_{n-2}$.

\end{document}